\documentclass[conference]{IEEEtran}

\usepackage{marvosym}

\usepackage{caption}
\usepackage{algpseudocode}
\usepackage{amsmath}
\usepackage{amsfonts}
\usepackage{amssymb}
\usepackage{subfigure}
\usepackage{graphicx}

\usepackage{amsfonts}
\usepackage{booktabs}
\usepackage{siunitx}

\usepackage[multiple]{footmisc}

\usepackage{color}

\usepackage{tikz}

\usepackage{enumitem}

\usepackage{url}
\makeatletter
\def\url@leostyle{%
  \@ifundefined{selectfont}{\def\UrlFont{\sf}}{\def\UrlFont{\small\ttfamily}}}
\makeatother
\newcommand{\eat}[1]{}
\usepackage[linesnumbered,ruled]{algorithm2e}
\usepackage{mdframed} 

\usepackage{xcolor}
\definecolor{light-gray}{gray}{0.9}

\newenvironment{packed_enum}{%
  \begin{enumerate}%
  }{\end{enumerate}}

\usepackage{amsthm}

\newcolumntype{L}[1]{>{\raggedright\let\newline\\\arraybackslash\hspace{0pt}}m{#1}}

\IEEEoverridecommandlockouts
\begin{document}
%-------------------------------------------------------------------------------

\title{Cross-Consensus Measurement of Individual-level Decentralization in Blockchains\\
% \thanks{This work was partially supported by the Beijing Natural Science Foundation under Grant No. M22039 and 4212008, the National Natural Science Foundation of China under Grant No. 62202038 and 62272031, and the Fundamental Research Funds for the Central Universities of China under Grant No. 2022JBMC007. 
% Runhua Xu is the corresponding author (runhua@buaa.edu.cn).}
}

\author{
  \IEEEauthorblockN{
    Chao~Li\IEEEauthorrefmark{1},
    Balaji~Palanisamy\IEEEauthorrefmark{2},
    Runhua~Xu\IEEEauthorrefmark{3} and
    Li~Duan\IEEEauthorrefmark{1}
  }
  \IEEEauthorblockA{
    \IEEEauthorrefmark{1}School of Computer and Information Technology, Beijing Jiaotong University, China \\
    % \IEEEauthorrefmark{2}Beijing Key Laboratory of Security and Privacy in Intelligent Transportation, Beijing Jiaotong University, China \\
    \IEEEauthorrefmark{2}School of Computing and Information, University of Pittsburgh, USA \\
    \IEEEauthorrefmark{3}School of Computer Science and Engineering, Beihang University, China \\
    li.chao@bjtu.edu.cn, bpalan@pitt.edu, runhua@buaa.edu.cn, duanli@bjtu.edu.cn
  }
}

\maketitle

%-------------------------------------------------------------------------------
\begin{abstract}
%-------------------------------------------------------------------------------
% Decentralization has been widely acknowledged as a core virtue of blockchain.
% , ignoring the underlying distribution of individual power that contributes to block producers.
Decentralization is widely recognized as a crucial characteristic of blockchains that enables them to resist malicious attacks such as the 51\% attack and the takeover attack.
Prior research has primarily examined decentralization in blockchains employing the same consensus protocol or at the level of block producers.
This paper presents the first individual-level measurement study comparing the decentralization of blockchains employing different consensus protocols.
To facilitate cross-consensus evaluation, we present a two-level comparison framework and a new metric.
We apply the proposed methods to Ethereum and Steem, two representative blockchains for which decentralization has garnered considerable interest.
Our findings dive deeper into the level of decentralization, suggest the existence of centralization risk at the individual level in Steem, and provide novel insights into the cross-consensus comparison of decentralization in blockchains.

\end{abstract}

\begin{IEEEkeywords}
  Blockchain, Decentralization, PoW, DPoS, Measurement Study.
\end{IEEEkeywords}

%-------------------------------------------------------------------------------
\section{Introduction}
%-------------------------------------------------------------------------------
Advances in blockchain technologies are fueling the emergence of various consensus protocols, including the Proof-of-Work (PoW) consensus protocol~\cite{nakamoto2008Bitcoin,buterin2014next}
% employed by Bitcoin~\cite{nakamoto2008Bitcoin} and Ethereum~\cite{buterin2014next}, as well as 
and the Delegated Proof-of-Stake (DPoS) consensus protocol~\cite{larimer2014delegated}.
%  that is gaining popularity. 
The PoW consensus protocol requires network-wide consensus to be reached across thousands of nodes, making it challenging to support high transaction throughput.
In contrast, the DPoS consensus protocol only requires a small committee of dozens of members to establish consensus, allowing DPoS blockchains to suit the practical needs of a broad range of decentralized applications.
In a DPoS blockchain, committee members generate blocks in rotation and jointly make decisions such as updating global parameters and restricting specific accounts.
Periodically, committee members are elected by coin holders who acquire and hold blockchain-issued coins (i.e., cryptocurrencies) and vote for their chosen candidates.
While the effectiveness of the underlying coin-based voting system has spawned successful DPoS blockchains such as Steem~\cite{guidi2020blockchain,li2021steemops} and EOSIO~\cite{zheng2021xblock}, the limited number of committee members in DPoS blockchains has raised concerns regarding centralization risk.

% decentralized applications, where authority and power are distributed across the network without the intervention of a single entity.
% The traditional Proof-of-Work (PoW) consensus protocol employed by Bitcoin~\cite{nakamoto2008Bitcoin} and Ethereum~\cite{buterin2014next} necessitates that the decentralized consensus is reached throughout the whole network.
% Consequently, the scale of the network restricts the throughput of transactions (e.g., 7 transactions per second in Bitcoin~\cite{croman2016scaling}), making it challenging to meet the requirements of many decentralized applications in practice.
% In order to address scalability concerns, recently, the Delegated Proof-of-Stake (DPoS) consensus protocol~\cite{larimer2014delegated} is gaining popularity and has spawned some successful blockchains, including Steem~\cite{guidi2020blockchain} and EOSIO~\cite{he2021eosafe}. 
% In a DPoS blockchain, a small group of block producers generates blocks on a rotating basis.
% Periodically, block producers are elected by coin holders, who acquire and hold coins (i.e., cryptocurrencies) issued by the blockchain and vote for their preferred candidates via a coin-based voting system.
% Therefore, the decentralized consensus in a DPoS blockchain is only achieved among a small group of block producers within a committee, allowing decentralized applications to enable high transaction throughput.

Decentralization is widely recognized as a crucial characteristic of blockchains that enables them to resist malicious attacks such as the 51\% attack~\cite{badertscher2021rational} for PoW blockchains and the takeover attack~\cite{ba2022fork} for DPoS blockchains.
For most public blockchains, the distribution of resources that determine who generates blocks is the key metric for evaluating decentralization~\cite{gencer2018decentralization,kwon2019impossibility,zeng2021characterizing}. This in turn facilitates further understanding of both security and scalability in blockchain~\cite{kokoris2018omniledger}.
Intuitively, having only a few parties possess the majority of the resources suggests a more centralized control of a blockchain, which is potentially less secure.
This is because the collusion of these few parties could jeopardize the immutability of blockchains by enabling the falsification of historical data that should have been confirmed.

Over the past few years, 
there has been an ongoing debate in the blockchain community with regard to the degree of decentralization in DPoS and PoW blockchains, resulting in two opposite positions:

\begin{itemize}[leftmargin=*]
\item Position I (PoW is more decentralized~\cite{Bitcoin_Wiki}): 
Theoretically, compared to the vast number of miners in PoW blockchains, the extremely small number of committee members in DPoS blockchains indicates a lower degree of decentralization.
\item Position II (DPoS is more decentralized~\cite{snider2018delegated}): 
In practice, recent research has demonstrated that Bitcoin and Ethereum have a tendency towards centralization~\cite{gencer2018decentralization,kwon2019impossibility}, therefore the design of rotating block production across committee members makes DPoS blockchains less centralized than the current PoW blockchains, which are dominated by a few mining pools.
% thus the design of equally generating blocks among committee memberes in DPoS blockchains is less centralized than the current PoW blockchains dominated by a few mining pools.
\end{itemize}

% Our results dive deeper into the level of decentralization DPoS blockchain achieves on the web, corroborate the existence of centralization risk and provide insights into the comparison of decentralization across DPoS and PoW blockchains.

% However, most of the existing measurement studies were collecting blocks generated in a single week and then computing the degree of decentralization based on the distribution of block producers within the selected week.
% Hence, their results could only reveal a snapshot state of the degree of decentralization, rather than its continuous shifts.
% Besides, some of these studies performed measurements using
% a single metric as well as a single granularity, making the
% results lack comprehensiveness.

% In this paper, we presents the first individual-level measurement study comparing the decentralization of Proof-of-Work (PoW) and Delegated Proof-of-Stake (DPoS) blockchains.
% To facilitate cross-consensus evaluation, we present a two-level measurement model and a new metric.
% We also propose a new algorithm for quantifying decentralization at the individual level in DPoS blockchains.
% We apply the proposed method to Ethereum and Steem, two representative blockchains for which decentralization has garnered considerable interest.

% In this paper, we present the first large-scale longitudinal study of the decentralization in Steem, a prominent DPoS blockchain that has served over one million social media users since 2016.
In this paper, we present a large-scale longitudinal measurement study comparing the decentralization of PoW and DPoS blockchains.
Our goals are twofold: 
to establish a finer-grained framework for cross-consensus comparisons of decentralization in blockchains,
and to develop a common methodology for gaining a deeper understanding of decentralization in DPoS blockchains.
% Besides, we apply the proposed method to Ethereum and Steem, two representative blockchains for which decentralization has garnered considerable interest~\cite{zeng2021characterizing,li2020comparison,li2019incentivized,guidi2020blockchain,jeong2020centralized}.

% to gain a deeper understanding of decentralization in DPoS blockchains, 
% and to provide novel insights into the comparison of decentralization at the individual level between DPoS and PoW blockchains
First, we notice that existing comparisons of decentralization between DPoS and PoW lack a multi-level framework capable of differentiating measurement granularity.
For instance, Position II compares the committee-level decentralization in DPoS with the pool-level decentralization in PoW.
In contrast, Position I compares the same committee-level decentralization in DPoS with the finer-grained pool-participant-level decentralization in PoW.
In this paper, to facilitate cross-consensus evaluation, we propose a two-level comparison framework for decentralization between DPoS and PoW blockchains that can distinguish and match measurements of different granularities.
Concretely, the proposed framework divides decentralization in DPoS into committee level and coin-holder level, and decentralization in PoW into pool level and pool-participant level.
The framework then benchmarks the committee level and coin-holder level in DPoS against the pool level and pool-participant level in PoW, respectively.
This is due to the fact that both DPoS and PoW blockchains are witnessing the pooling of resources from individuals (coin holders and pool participants) to representatives (committee members and mining pools), leading to two different levels of resource concentration.
% the two opposite positions compare committee-level decentralization in DPoS with pool-level (Position II) and participant-level (Position I) decentralization in PoW, respectively.
% In this paper, we compare decentralization across DPoS and PoW using a two-level reference model shown in Figure~\ref{sec1_01}, where committee member level and shareholder level in DPoS are benchmarked against mining pool level and pool participant level in PoW, respectively.
% This is due to the fact that, despite their different types of resources determining block producers, both DPoS and PoW blockchains are committee membering pooling of resources from individuals (coin holders and participants) to pools (committee memberes and mining pools), resulting in two different levels of concentration of resources.
% Our study concludes that, compared with Ethereum, Steem tends to be more decentralized at the pool level but less decentralized at the individual level.

Second, existing studies on blockchain decentralization mainly focus on the decentralization of mining power in PoW blockchains~\cite{gencer2018decentralization,gervais2014bitcoin,zeng2021characterizing}. 
Among them, the state-of-the-art study suggests that decentralization assessed at a deeper level (i.e., across participants of mining pools) could be substantially distinct from that measured across mining pools~\cite{zeng2021characterizing}.
Inspired by prior research, we argue that both of the two opposite positions (I\&II) mainly focus on decentralization at the level of representatives, namely mining pools in PoW and committee members in DPoS.
Neither of these positions goes deeper into the finer-grained level of individuals.
To obtain a more comprehensive understanding of decentralization in blockchains, our analysis is performed at two different levels, the representative level, where committee members in DPoS and mining pools in PoW produce blocks, and the deeper individual level, where coin holders contribute voting power to committee members in DPoS and pool participants contribute mining power to mining pools in PoW.
% However, diving deeper into the individual level is non-trivial, especially for DPoS blockchains.
% In this paper, we carefully investigate the underlying coin-based voting system of DPoS and propose a new algorithm, DPoS-ILDQ, for quantifying decentralization at the individual level for DPoS blockchains.

We apply the proposed framework and methodologies to Ethereum\footnote[1]{The Ethereum community has recently launched Ethereum 2.0 that adopts Proof-of-stake (PoS) in Sep. 2022. 
Following recent works on measurement study of decentralization in Ethereum, 
this paper focuses on a four-year period of time of Ethereum 1.0 that employs PoW. 
In this rest of this paper, unless explicitly stated, we use Ethereum to refer to Ethereum 1.0.} and Steem, two representative blockchains for which decentralization has garnered considerable interest~\cite{gencer2018decentralization,kwon2019impossibility,li2020comparison,zeng2021characterizing}.
In Ethereum, inspired by prior research, we estimate the mining power possessed by a pool participant based on the amount of reward the participant received from mining pools.
In Steem, we reconstruct the historical election snapshots on a daily basis for a period of four years.
Our study suggests that, based on the proposed methods, Steem tends to be less decentralized than Ethereum at the individual level and more vulnerable to malicious attacks such as takeovers.
% Overall, our results dive deeper into the level of decentralization, suggest the existence of centralization risk at the individual level in Steem, and provide novel insights into the comparison of decentralization between PoW and DPoS blockchains.

% To facilitate measurements from different perspectives, we propose a measurement algorithm DPoS-ILDQ and two new metrics, the top-$k$ normalized entropy (NE) coefficient and the minimum threshold (MT) coefficient.
% Besides, to dive deeper into the individual level in Steem, we fix missing system parameters and reconstruct the historical stake snapshots on a daily basis for a period of five years.
% Our study concludes that, before the takeover, Steem tends to be less decentralized than Ethereum at the individual level.

% Overall, our results dive deeper into the level of decentralization DPoS blockchain achieves on the web, corroborate the existence of centralization risk and provide novel insights into the comparison of decentralization across DPoS and PoW blockchains.

\noindent \textbf{Contributions }
In a nutshell, this paper makes the following key contributions:

\begin{itemize}[leftmargin=*]
\item To the best of our knowledge, our work presents the first individual-level measurement study comparing the decentralization of PoW and DPoS blockchains.
\item Our work dives deeper into the level of decentralization, suggests the existence of centralization risk at the individual level in Steem. and provides novel insights into the comparison of decentralization between PoW and DPoS blockchains.
% \item We believe that the individual-level decentralization of any blockchain other than Steem and Ethereum remains unstudied. Therefore, we believe the methods proposed in this work, including the double perspective on decentralization and the methodology and metrics, could facilitate future works on drawing a more generic comparison over different consensus protocols.
% \item Our work provides novel insights into the correlation between decentralization and takeover.
% Our study suggests that the targeted blockchain of the first de facto takeover is relatively less decentralized than the well-studied Ethereum blockchain.
% Our study also demonstrates the long-term damage to the decentralization of both Steem and Hive after the takeover, which may suggest that no one won the battle from this perspective.
% \item Our work corroborates the existence of the Sword of Damocles over DPoS. That is, the neutral actors (e.g., accounts holding pre-mined tokens, exchanges holding users' tokens) may one day collude to defeat the rest users and take over the blockchain. 
\end{itemize}

\begin{table}[tp]  
  \centering   
  \caption{Comparison of Emerging Studies on Decentralization in Blockchains}  
  \label{tab:33} 
\begin{tabular} 
%{cccccc}  
{p{2.3cm}<{\centering} p{2.0cm}<{\centering} p{2.1cm}<{\centering} }
\hline
\hline 
{\bf Related work} & {\bf Cross-consensus} & {{\bf Individual-level}}\\ 
\hline
Gervais et al.~\cite{gervais2014bitcoin} & $\times$  & $\times$ \\
Miller et al.~\cite{miller2015discovering} & $\times$  & $\times$ \\
Gencer et al.~\cite{gencer2018decentralization} & $\times$  & $\times$ \\
Lin et al.~\cite{lin2021measuring} & $\times$  & $\times$ \\
Zeng et al.~\cite{zeng2021characterizing} & $\times$  & $\surd$ \\
Kwon et al.~\cite{kwon2019impossibility} & $\surd$  & $\times$ \\
Li et al.~\cite{li2020comparison} & $\surd$  & $\times$ \\
Our work & $\surd$  & $\surd$ \\
\hline
\hline 
\end{tabular}
%\vspace{-7mm}
\end{table}

\section{Related work}
\label{Related}
% \noindent \textbf{Evaluation of decentralization.}
Over the past few years, many studies have measured and evaluated the decentralization of the two major PoW blockchains, Bitcoin and Ethereum, from multiple perspectives.
In~\cite{gervais2014bitcoin}, Gervais et al. showed that some fundamental system modules in Bitcoin, such as decision-making and mining,  were not decentralized. 
They showed that a limited number of entities dominated these system modules in Bitcoin.
Later, in~\cite{miller2015discovering}, Miller et al. developed AddressProbe, a tool to discover links in the underlying peer-to-peer network of Bitcoin and construct the live topology.
Based on the measured topology, their results found that, in contrast to the idealistic vision of distributing mining power across nodes, several prevalent and hidden mining pools were controlling the majority of mining power.
In 2018, Gencer et al. provided a measurement analysis of a variety of decentralization metrics for Bitcoin and Ethereum~\cite{gencer2018decentralization}. Their research evaluated the network resources of nodes and their connectivity, as well as the attack resilience of the two blockchains.
Their results suggested that there was no significant difference between decentralization in Bitcoin and in Ethereum.
In 2021, with different metrics and granularities, Lin et al. revealed that the degree of decentralization in Bitcoin is higher, while the degree
of decentralization in Ethereum is more stable~\cite{lin2021measuring}.
The state-of-the-art study~\cite{zeng2021characterizing} measured the decentralization in Ethereum at the level of mining pool participants. The results indicated that decentralization measured at a deeper level could be quite different from that measured across mining pools.

With the rapid development of DPoS blockchains, several recent studies have evaluated the decentralization in DPoS blockchains.
In~\cite{kwon2019impossibility}, Kwon et al. quantified the decentralization in dozens of PoW, PoS and DPoS blockchains. Their work helps build a broad understanding of the representative-level decentralization of blockchains adopting different consensus protocols.
However, their results may have some limitations because they only collected blocks generated in Oct. 2018.
Later, in~\cite{li2020comparison}, Li et al. leveraged the Shannon entropy to quantify the decentralization in Steem and Bitcoin. Their results revealed that, compared with Steem, Bitcoin tends to be more decentralized among top miners but less decentralized in general. 
However, without reconstructing the historical election snapshots, their study only measured a one-month snapshot of decentralization in Steem.

To sum up, as illustrated in TABLE~\ref{tab:33}, to the best of our knowledge, we believe that our work presents the first cross-consensus measurement study of individual-level decentralization in Blockchains.

\section{Background}
\label{Background}

In this section, we introduce the background about the Ethereum blockchain~\cite{wood2014ethereum} and the Steem blockchain~\cite{Steem_blockchain}, including their ecosystem in general, their implementation of PoW and DPoS consensus protocols, the main attacks they face and the correlation between these attacks and the degree of decentralization.

% its most popular decentralized application \textit{Steemit}, its implementation of the DPoS consensus protocol and its ecosystem in general.

\subsection{Ethereum}

Ethereum is the representative of the second-generation blockchain, and its market capitalization is perennially in the top two.
The underlying Ethereum Virtual Machine (EVM) enables the development of powerful smart contracts and has supported thousands of decentralized applications, ranging from exchanges to games.
Besides, Ethereum issues a cryptocurrency named Ether (ETH) to support the establishment of its decentralized financial ecosystem.

\noindent \textbf{Consensus protocol in Ethereum 1.0.}
In Ethereum, a user needs to create an account to interact with the blockchain.
Users may perform three categories of transactions, namely transferring funds, creating smart contracts and invoking functions within deployed smart contracts.
A user needs to build a transaction locally and transfer it to Ethereum's underlying peer-to-peer (P2P) network for the transaction to be executed.
When miners within the P2P network receive a transaction, they cache it in a pool.
In the meantime, miners continue to work on a mathematical puzzle.
After solving the puzzle, miners package the cached transactions into a block.
The block is then broadcast, validated by other miners and finally attached to the blockchain.

\noindent \textbf{51\% attack and decentralization.}
In PoW blockchains such as Bitcoin and Ethereum, mining power refers to a miner's ability to solve the mathematical puzzle. 
If a single miner's mining power is larger than the total of all other miners' mining power, that miner will be able to control the production of blocks and falsify confirmed transactions, such as double spending cryptocurrency.
Consequently, the decentralization of mining power in Ethereum has recently garnered a great deal of interest.~\cite{gencer2018decentralization,zeng2021characterizing}.

\subsection{Steem}
\begin{figure}
\centering
{
    \includegraphics[width=0.9\columnwidth]{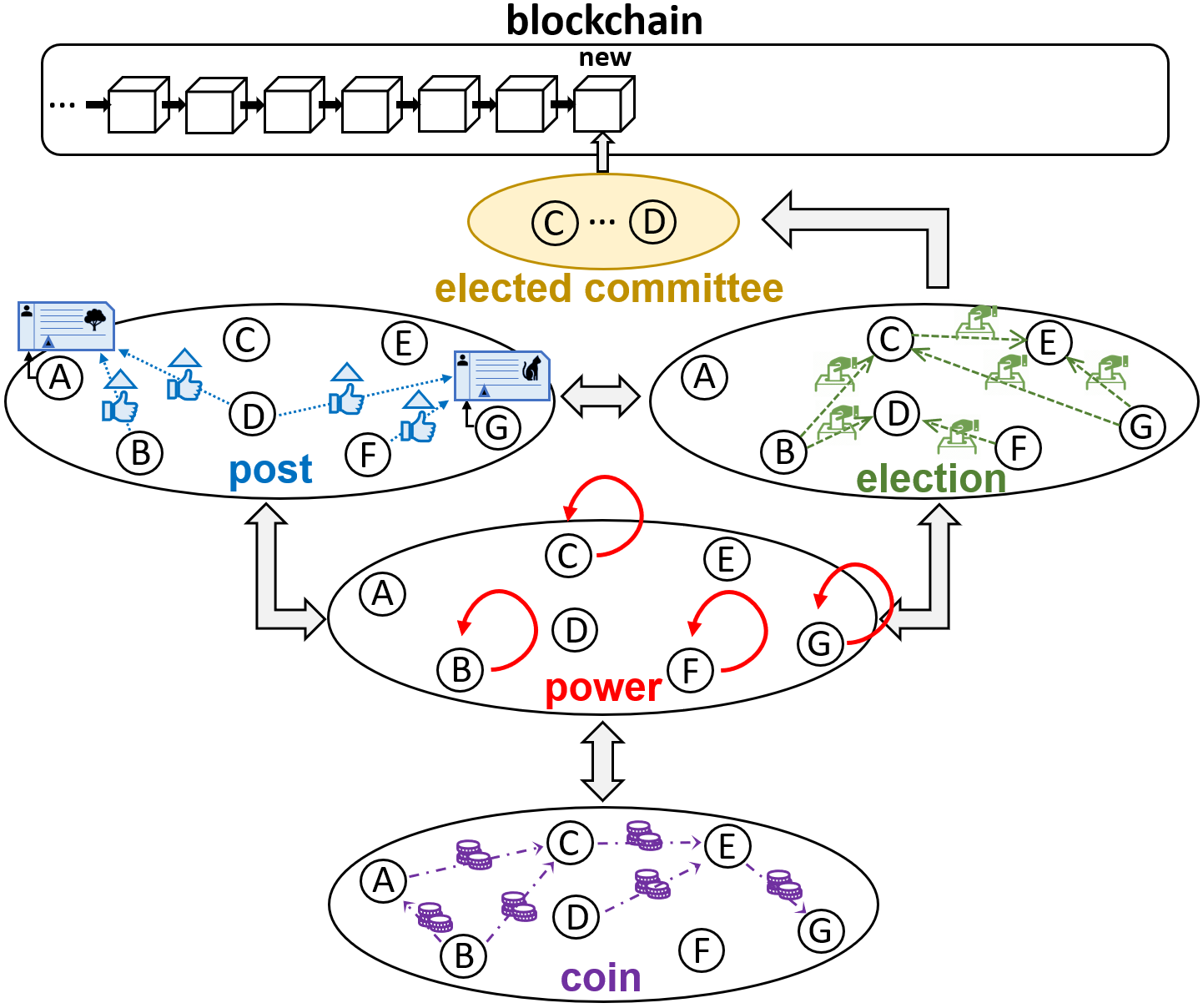}
}
% \vspace{-3mm}
\caption {Steem blockchain overview}
\vspace{-3mm}
\label{steem_overview} 
\end{figure}

Steem serves over one million users and has recorded around one billion operations.
Similar to Ethereum, Steem is a blockchain built to serve a wide variety of applications.
For instance, the most popular application, named \textit{Steemit}~\cite{li2019incentivized}, is comparable to a decentralized \textit{Reddit}, where users can upvote and downvote blog posts published by other users. 
There have been around thirty different sorts of operations in Steem.
In Figure~\ref{steem_overview}, we present four representative categories of operations, namely social post operations, committee election operations, voting power operations and coin transfer operations.
Like most blockchains, Steem issues a cryptocurrency known as \textit{STEEM}.

\noindent \textbf{Consensus protocol in Steem.}
On the basis of DPoS, Steem encourages coin holders to cast up to 30 votes for committee candidates.
The top 20 candidates then form a committee charged with maintaining high-quality servers that generate blocks periodically.
Due to the small size of the committee, the transaction throughput could be greatly enhanced.
In committee elections, coin holders' votes are weighted according to the voting power derived from their coins.
Specifically, coin holders must freeze coins (i.e., \textit{STEEM}) to obtain voting power so that they can support the candidates of their choice at the expense of freezing the liquidity of their coins.
Coin holders may unfreeze their frozen coins at any time, but they will immediately lose the corresponding amount of voting power, and the frozen coins will be automatically divided into thirteen parts and withdrawn in thirteen weeks.
In addition to personally voting for candidates, a coin holder may appoint another coin holder as her proxy to vote on her behalf.
The use of proxies complicates the reconstruction of historical election snapshots, as the proxy appointed by a coin holder may have also appointed a proxy, leading to a proxy chain.

\noindent \textbf{Takeover attack and Decentralization.}
In DPoS blockchains such as Steem~\cite{guidi2020blockchain} and EOSIO~\cite{zheng2021xblock}, committees undertake on-chain governance in addition to producing blocks.
Specifically, committee members may decide on proposals such as updating blockchain parameters and even banning certain accounts.
A proposal must have at least 17 (15) approvals from Steem (EOSIO) committee members in order to be implemented.
After receiving sufficient approvals, a proposal takes immediate effect at the code level.
As can be seen, if a single coin holder possesses a significant amount of voting power, the coin holder may be able to use the 30 votes to directly determine the 20 committee seats and gain entire control of the committee with 20 controlled accounts.
This powerful coin holder could then implement any proposals, such as banning accounts against such a takeover or even reversing confirmed transactions.
Steem was the victim of the first de facto blockchain takeover.
On March 2, 2020, all Steem committee members were suddenly replaced with those controlled by a single coin holder, who then immediately blacklisted some of the original committee members, causing the Steem community to split.
Consequently, the decentralization of committee election in Steem has garnered considerable attention~\cite{guidi2020blockchain,jeong2020centralized,li2020comparison}.

Next, based on the background introduced in this section, we present a two-level comparison framework and a new metric in Section~\ref{Model and Metric}.

\begin{figure}
  \centering
  {
      \includegraphics[width=0.89\columnwidth]{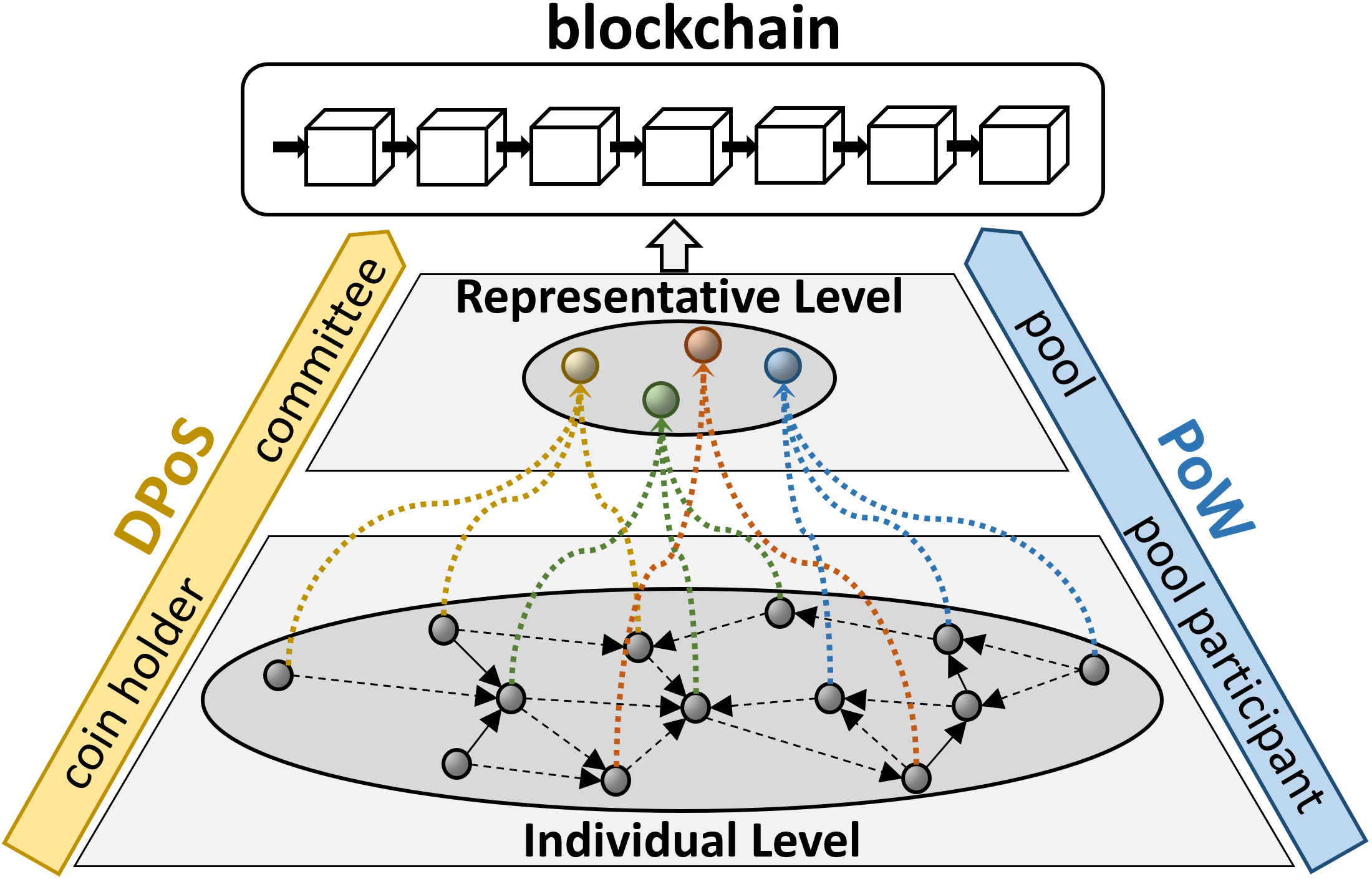}
  }
  % \vspace{-3mm}
  \caption {A two-level framework for comparing decentralization between DPoS and PoW blockchains}
  \vspace{-3mm}
  \label{model} 
\end{figure}

\section{Comparison Framework and Metric}
\label{Model and Metric}
In this section, we present a two-level framework and also a new metric for comparing decentralization between DPoS and PoW blockchains. 

\subsection{Two-level comparison framework}

% \noindent \textbf{Two-level measurement model.}
In the absence of a multi-level framework capable of discriminating measurement granularity, we propose a two-level comparison framework that can differentiate and match measurements of different granularities.
The framework is shown in Fig.~\ref{model}.
The design of the framework is inspired by the observation that both DPoS and PoW blockchains include two levels of resource concentration as a result of the pooling of resources from individuals to representatives.

\noindent \textbf{Representative level.}
In PoW blockchains, almost all blocks are now produced by mining pools.
In DPoS blockchains, only committee members are permitted to generate blocks.
Together, we can see that in both DPoS and PoW blockchains, despite the difference in consensus protocol, a small number of entities hold the rights to produce blocks.
In this paper, mining pools and committee members are considered to be representatives of pool participants and coin holders, respectively.
Concretely, mining pools in PoW acquire mining power from individual pool participants to compete with each other.
Similarly, candidates in DPoS request voting power from individual coin holders to enter the top 20.
In other words, pool participants and coin holders utilize their individual power to select representatives who will generate blocks on their behalf.
The difference is that the selection of representatives is voluntary in PoW but mandatory in DPoS.
Therefore, the framework benchmarks the committee-level decentralization in DPoS against the pool-level decentralization in PoW.
Specifically, committee-level decentralization in DPoS could be assessed by analyzing the distribution of the number of blocks produced by each committee member.
Similarly, pool-level decentralization in PoW refers to the distribution of the number of blocks produced by different mining pools. 

\noindent \textbf{Individual level.}
Most prior discussions and research have focused on the decentralization of blockchains at the representative level.
In this paper, we suggest that the root causes of the 51\% attack and the takeover attack should be the individuals, not the representatives.
First, representatives have no control over the power granted to them by individuals.
For instance, if several representatives plan to conclude and launch an attack, individuals would be able to withdraw their power from these representatives to stop the attack.
Second, an individual may control multiple representatives.
Recall that during the Steem takeover, a single coin holder controlled the whole committee.
This coin holder's dominance may be observed at the individual level but not at the representative level.
Therefore, this paper focuses primarily on decentralization at the individual level.
The proposed framework benchmarks the coin-holder-level decentralization in DPoS against the pool-participant-level decentralization in PoW.
% Next, we propose a new metric for measuring and comparing individual-level decentralization.
% Concretely, in PoW blockchains such as Ethereum, we estimate the mining power possessed by a pool participant based on the amount of total reward the participant received from all mining pools.

% First, we notice that existing comparisons of decentralization between DPoS and PoW lacks a multi-level framework capable of differentiating measurement granularity.
% In this paper, to facilitate cross-consensus evaluation, we propose a two-level comparison framework for decentralization between DPoS and PoW blockchains that can distinguish and match measurements of different granularities.
% Concretely, the proposed framework divides decentralization in DPoS into committee level and coin-holder level, and decentralization in PoW into pool level and pool-participant level.
% The framework then benchmark committee level and coin-holder level in DPoS against pool level and pool-participant level in PoW, respectively.
% This is due to the fact that both DPoS and PoW blockchains are committee membering pooling of resources from individuals (coin holders and pool participants) to representatives (committee members and mining pools), leading in two different levels of resource concentration.

\subsection{Metric}
Before presenting the proposed metric, we will first define the term ``individual impact''.

\noindent \textbf{Individual impact.}
In this paper, we estimate individual impacts in block-production competitions based on a data-driven approach.
Specifically, in Ethereum, we estimate the per-day impact of a pool participant based on the total rewards received by the participant from all mining pools.
In Steem, we first extract blocks produced by committee members and then allocate these blocks to coin holders based on daily election snapshots.
That is, the impact of a particular coin holder refers to the sum of blocks allocated from all committee members who have received votes from either the coin holder or the coin holder's proxy.
In the rest of this paper, we will denote individuals as a $n$-dimensional vector $\mathbf{s}$, and impacts of individuals as a $n \times m$ matrix $\mathbf{A}$, where the entry $a_{ij}$ denotes the impact of $i^{th}$ individual on $j^{th}$ day.

\iffalse
\noindent \textbf{Top-$k$ Normalized Entropy (NE) coefficient.}
Whales are individuals with enormous resources and are usually the most influential players in block-creation competitions.
Therefore, an important perspective of decentralization is the evolution of whales and the distribution of impacts among them.
We propose the top-$k$ NE coefficient to quantify entropy-based decentralization among all time top-$k$ whales 
on a daily basis by computing normalized entropy for $j^{th}$ day as:
\begin{equation}
e_j = - \sum_{i=1}^{o_j} \frac{p_{ij} \log_2 p_{ij}}{\log_2 o_j} \label{e2}
\end{equation}
where $o_j$ denotes the amount of active top-$k$ whales with non-zero impacts on $j^{th}$ day, 
and $p_{ij}$ denotes the percentage of $i^{th}$ whale's impact among active top-$k$ whales on $j^{th}$ day.
\fi

% A similar metric, called Nakamoto coefficient~\cite{fritsch2022analyzing,lin2021measuring}, has been widely used in evaluating the decentralization of blockchains, which basically measures the minimum number of holders of a certain type of resource (e.g., mining power) such that the sum of their resources is larger than a threshold (e.g., 50\%).

\noindent \textbf{Minimum Threshold (MT) coefficient.}
Previous research has extensively employed the Nakamoto coefficient~\cite{nc} to quantify the decentralization of mining power in PoW blockchains.
The Nakamoto coefficient is defined as the minimum number of mining pools whose combined mining power exceeds fifty percent of the total mining power of the entire blockchain.
In this paper, inspired by the Nakamoto coefficient, we propose the Minimum Threshold (MT) coefficient to quantify daily threshold-based decentralization by computing the minimum number of individuals whose combined power exceeds a particular threshold as:
\begin{equation}
  \label{e1}
f_j = min(\mathbf{s},> t \cdot sum(\mathbf{A}_{*,j}))
\end{equation}
where $sum(\mathbf{A}_{*,j})$ represents the sum of individual impacts on $j^{th}$ day, $t$ represents the threshold,
and $min(\mathbf{s},condition)$ represents the minimum number of individuals whose combined power meets the condition.
The suggested metric is compatible with both DPoS and PoW blockchains, thus facilitating cross-consensus comparisons.

\section{Data collection}
\label{Data collection}
In this section, we outline our process for collecting data.
The Steem blockchain provides developers and researchers with an Interactive Application Programming Interface (API) at \cite{SteemAPI} to collect and parse blockchain data.
To analyze the decentralization of Steem prior to the takeover on 2020-03-02, we collected over 41 million blocks generated between Steem's birthday (2016-03-24) and the day of the takeover (2020-03-02).
% In TABLE~\ref{t1}, we describe six types of Steem operations that are crucial to this study.
For Ethereum, since we are only interested in its degree of decentralization before the Steem takeover on 2020-03-02, we collected block data from 2016-03 to 2020-02 using the Ethereum official API at~\cite{EtherscanAPI}.
To explore decentralization at the individual level, we extract the Ethereum participant reward data (i.e., rewards paid from mining pools to pool participants) from the dataset published by the recent work~\cite{zeng2021characterizing}.

\section{Measuring Decentralization}
\label{Measuring}
In this section, we measure and compare both the representative-level decentralization and the individual-level decentralization between Steem and Ethereum.

\begin{figure}
\centering
{
    \includegraphics[width=0.99\columnwidth]{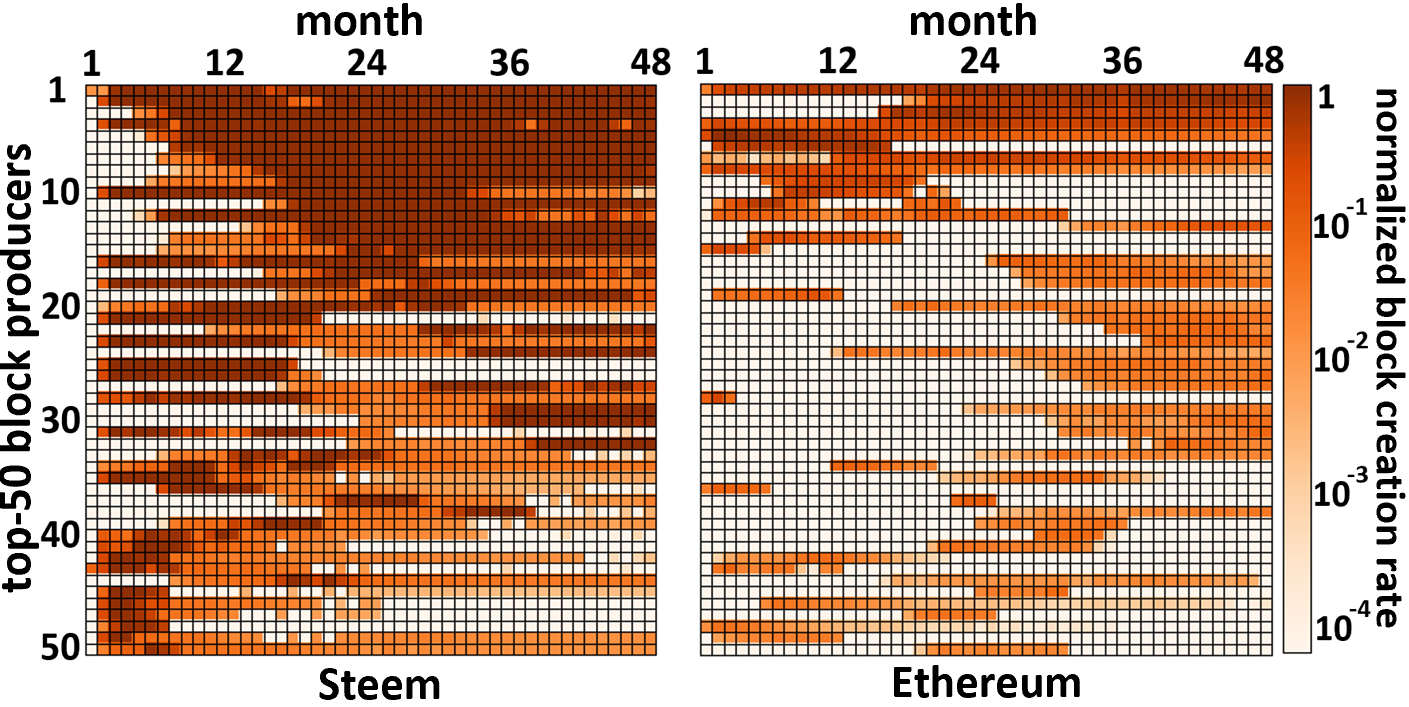}
}
\vspace{-2mm}
\caption {Heatmap of top-50 producers' normalized block creation rates during 48 months in Steem and Ethereum}
\vspace{-3mm}
\label{heatmap_1} 
\end{figure}

% \subsection{Visualization methodology}
% \label{visual}
% There are four key steps:
% % \begin{mdframed}[innerleftmargin=1.6pt]
%   \begin{packed_enum}[leftmargin=*]
%     \item Extract block producers (representatives) from raw data.
%     \item Calculate the number of blocks produced per producer per month, order the producers by the total number of blocks produced by each producer during the 48 months from 2016-03 to 2020-02, and identify the top-$l$ producers of all time.
%     \item For the top-$l$ producers, create a $l \times 48$ matrix $\mathbf{B}$, where the entry $b_{ij}$ represents the number of blocks produced by the $i^{th}$ top-$l$ producer in the $j^{th}$ month.
%     Then, compute the normalized block creation rates as a $l \times 48$ matrix $\mathbf{C}$, where
%     \begin{equation}
%       c_{ij} = \frac{b_{ij}}{max(\mathbf{B})}
%     \end{equation}
%     namely the ratio between the number of blocks created by the $i^{th}$ producer during the $j^{th}$ month and the maximum number of blocks produced by any producer in any month.
%     \item Finally, visualize the results as a heatmap, where the cell $(i,j)$ corresponds to the entry $c_{ij}$.
%   \end{packed_enum}
%   % \end{mdframed}

\subsection{Representative-level Decentralization}
\label{pool}

\noindent \textbf{Methodology.} 
There are four key steps.
First, we extract block producers (representatives) from raw data.
Then, we calculate the number of blocks produced per producer per month, order the producers by the total number of blocks produced by each producer during the 48 months from 2016-03 to 2020-02, and identify the top-$l$ producers of all time.
After that, for the top-$l$ producers, we create a $l \times 48$ matrix $\mathbf{B}$, where the entry $b_{ij}$ represents the number of blocks produced by the $i^{th}$ top-$l$ producer in the $j^{th}$ month. 
We compute the normalized block creation rates as a $l \times 48$ matrix $\mathbf{C}$, where
  \begin{equation}
    c_{ij} = \frac{b_{ij}}{max(\mathbf{B})}
  \end{equation}
  namely the ratio between the number of blocks created by the $i^{th}$ producer during the $j^{th}$ month and the maximum number of blocks produced by any producer in any month.
Finally, we visualize the results as a heatmap, where the cell $(i,j)$ corresponds to the entry $c_{ij}$.

\noindent \textbf{Results and insights.}
In Figure~\ref{heatmap_1}, we observe that Steem has much more $c_{ij}$ close to 1 than Ethereum. 
This is because the top Steem producers create blocks in a rotational manner, whereas the top Ethereum producers follow a winner-takes-all model.
Consequently, most Steem producers have their monthly produced blocks $b_{ij}$ close to the theoretical maximum\footnote[2]{In Steem, during a month that has thirty days, there should be $30*24*60*60/3=$ 864,000 blocks generated because blocks are generated every three seconds.
For every 21 blocks, the 21 elected committee members are shuffled to determine their order for generating the next 21 blocks, so a committee member can at most produce $864000/21 \approx$ 41,142 blocks in a thirty-day month.}, while the Ethereum mining competition is dominated by a few leading mining pools.
Next, by longitudinally observing the change of $c_{ij}$, we could find interesting similarities between Steem and Ethereum, such as the fact that first-tier producers do not always change, whereas second-tier producers have experienced at least two generations.
% Specifically, in Steem, from month 20 to month 48, the top-15 committee memberes firmly held at least 12 seats.
% In contrast, for the rest of the top-50 committee memberes, we can observe a transition period, namely month 20 to month 30, during which the places of old committee memberes were gradually taken by new committee memberes.
% A similar fact could be found in Ethereum, where top-8 producers create the majority of blocks, and old producers outside top-8 were gradually taken by new producers during month 20 to month 30.

% \noindent \textbf{Insights }
Overall, our results suggest that the Steem blockchain tends to be more decentralized than Ethereum at the representative level.
In other words, by looking only at the producers who directly create blocks, the blocks in Steem are more evenly distributed across producers.

\subsection{Individual-level Decentralization}
\label{individual}

\begin{figure}
  \centering
  {
      \includegraphics[width=1\columnwidth]{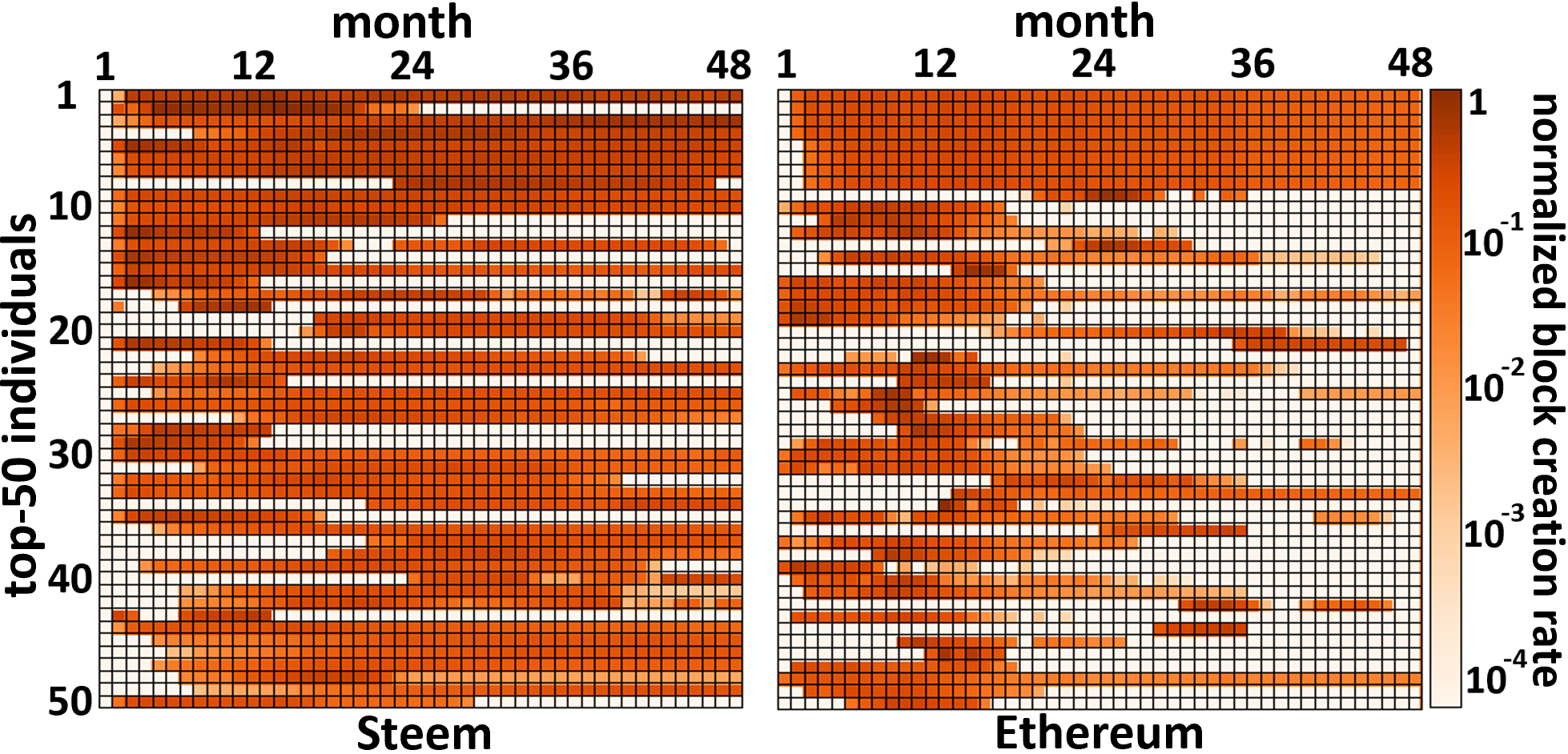}
  }
  % \vspace{-3mm}
  \caption {Heatmap of top-50 individuals' normalized block creation rates during 48 months in Steem and Ethereum}
  \vspace{-2mm}
  \label{heatmap_3} 
  \end{figure}

\noindent \textbf{Methodology.} 
  In Ethereum, we estimate the individual impact of a pool participant based on the amount of reward the participant received from all mining pools.
  In Steem, we estimate the individual impact of a coin holder based on the historical election snapshots reconstructed on a daily basis.
Then, based on the estimated individual impacts, we could allocate blocks produced by representatives to individuals and modify the methodology presented in Section~\ref{pool} by replacing blocks produced by committee members and mining pools with blocks allocated to coin holders and pool participants.

\noindent \textbf{Results and insights }
% Figure~\ref{sec5_3_03} indicates that, over time, the amount of active top-100 coin holders in Steem shows a growing trend, while the amount of active top-100 pool participants in Ethereum shows a downward trend.
% During month 36 to month 48, the amount of active top-100 individuals stabilize at around 85 and 20 in Steem and Ethereum, respectively.
% The change in the amount of active top-100 individuals suggests the use of normalized entropy.
% Given $n$, normalized entropy can normalize the value of entropy by the theoretical maximum $\log_2 n$.
Figure~\ref{heatmap_3} shows the heatmap measured using the above methodology.
% It replaces blocks produced by committee members and mining pools with blocks allocated to coin holders and pool participants.
We could see that the normalized block creation rates of top individuals are relatively lower than those of top producers displayed in Figure~\ref{heatmap_1}, either in Steem or Ethereum.
Compared with Ethereum, Steem tends to have more active top individuals, especially after month 20.

\iffalse
Figure~\ref{NE_1} shows the results of measurements for top-100 normalized entropy (NE) coefficient.
We can see that, during the early months 2 to 15, Ethereum always has a higher NE coefficient than Steem.
However, after month 15, the two blockchains are getting comparable.
Specifically, in Ethereum, its NE coefficient keeps dropping from month 2 to month 25. After a sudden surge, its NE coefficient stabilizes at around 0.85.
In Steem, its NE coefficient tends to be more stable and is also close to 0.85, on average.

% drops from month 4 to month 15, and then gradually increases from month 15 to month 33. However, from month 33 to month 48, its normalized entropy drops from 0.7 to nearly 0.6.

\begin{figure}
\centering
{
    \includegraphics[width=1.0\columnwidth]{./figures/usenix_entropy_before.eps}
}
\vspace{-3mm}
\caption {\small The top-100 NE coefficient computed on a daily basis in Steem and Ethereum during 48 months}
% \vspace{-2mm}
\label{NE_1} 
\end{figure}
\fi

\begin{figure}
  \centering
  {
      \includegraphics[width=0.9\columnwidth]{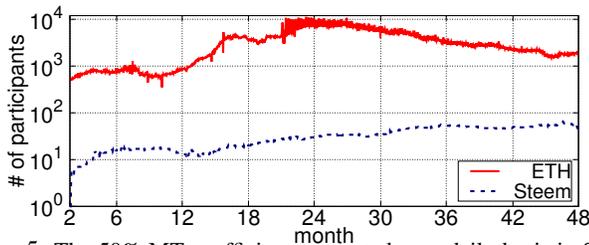}
  }
  \vspace{-3mm}
  \caption {\small The 50\% MT coefficient computed on a daily basis in Steem and Ethereum during 48 months}
  \vspace{-3mm}
  \label{MT_1} 
\end{figure}

Figure~\ref{MT_1} presents the results of measurements for the 50\% minimum threshold (MT) coefficient.
Compared to Ethereum, Steem has substantially lower MT coefficients over all months.
Specifically, the minimum, maximum, and average MT coefficients for Ethereum are 363, 1916, and 1140, respectively.
In comparison, the minimum, maximum and average MT coefficients for Steem are 1, 49 and 25, respectively.
Despite the vast disparity, we see that Steem's MT coefficient is gradually increasing while Ethereum's MT coefficient is slowly decreasing.

% \noindent \textbf{Insights }
Overall, the results suggest that the Steem blockchain tends to be less decentralized than Ethereum at the individual level.
That is, by diving deeper into the individual level, resources, such as accumulated voting power in Steem or mining power in Ethereum, tend to be dispersed more evenly in Ethereum.
The longitudinal observation reveals that Steem shows a significant disadvantage with regard to MF coefficients, which may suggest that Steem is more vulnerable to malicious attacks such as takeovers.
\section{Conclusion}
\label{Conclusion}

In this paper, we present the first double-level longitudinal measurement study comparing the decentralization of PoW and DPoS blockchains.
To facilitate cross-consensus comparison, we present a two-level comparison framework and a new metric named MT coefficient.
We apply the proposed methods to Ethereum and Steem.
Our results suggested that, compared with Ethereum, Steem tends to be more decentralized at the representative level but less decentralized at the individual level.
Our results also suggest that Steem may be more vulnerable to malicious attacks such as takeovers.
We believe the methods proposed in this work, including the double perspective on decentralization and the methodology and metrics, could facilitate future works on drawing a more generic comparison over different consensus protocols.

\renewcommand\refname{Reference}

\bibliographystyle{plain}
\urlstyle{same}

\bibliography{main.bib}

% \appendix

% \section{A}

% \begin{figure}
% \centering
% {
%     \includegraphics[width=1.0\columnwidth]{./figures/heatmap_afterfork.png}
% }
% \vspace{-6mm}
% \caption {Heatmap of top-50 producers' normalized block creation rates during 48 months in Steem and Ethereum}
% \vspace{-2mm}
% \label{sec3_01} 
% \end{figure}

% \begin{figure}
% \centering
% {
%     \includegraphics[width=1.0\columnwidth]{./figures/heatmap_afterfork_stakeholder.png}
% }
% \vspace{-6mm}
% \caption {Heatmap of top-50 producers' normalized block creation rates during 48 months in Steem and Ethereum}
% \vspace{-2mm}
% \label{sec3_01} 
% \end{figure}

\end{document}